\documentclass[aps,prl,reprint,groupedaddress,floatfix]{revtex4-1}

\usepackage{graphicx}
\usepackage{siunitx}

\begin{document}

\title{Low dimensionality of the diamond surface conductivity}

\author{Moritz V. Hauf, Patrick Simon, Max Seifert, Alexander W. Holleitner, Martin Stutzmann}
\affiliation{Walter Schottky Institute, Technische Universit{\"a}t M{\"u}nchen, Am Coulombwall 4, 85748 Garching, Germany}

\author{Jose A. Garrido}
\email{JoseAntonio.Garrido@wsi.tum.de}
\affiliation{Walter Schottky Institute, Technische Universit{\"a}t M{\"u}nchen, Am Coulombwall 4, 85748 Garching, Germany}

\date{\today}

\begin{abstract}
Undoped diamond, a remarkable bulk electrical insulator, exhibits a high surface conductivity in air when the surface is hydrogen-terminated. Although theoretical models have claimed that a two-dimensional hole gas is established as a result of surface energy band bending, no definitive experimental demonstration has been reported so far. Here, we prove the two-dimensional character of the surface conductivity by low temperature characterization of diamond in-plane gated field-effect transistors that enable the lateral confinement of the transistor's drain-source channel to nanometer dimensions. In these devices, we observe Coulomb blockade effects of multiple quantum islands varying in size with the gate voltage. The charging energy and thus the size of these zero-dimensional islands exhibits a gate voltage dependence which is the direct result of the two-dimensional character of the conductive channel formed at hydrogen-terminated diamond surfaces.
\end{abstract}


\maketitle

\begin{figure}
	\includegraphics{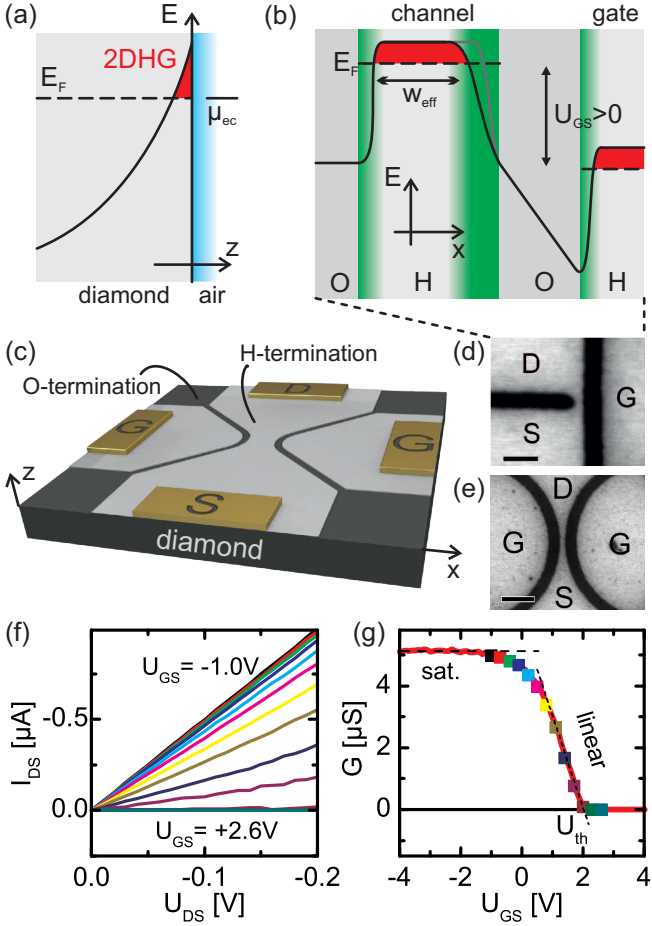}
	\caption[Basic principles and characterization of a diamond in-plane gated field effect transistor.]{\label{fig:1_IPGFET_RT} Basic principles and characterization of a diamond in-plane gated field effect transistor. (a) Energy band schematic of a H-terminated diamond surface. Equilibration of $E_F$ with $\mu_{EC}$ leads to a bending of the valence band edge and the creation of a 2DHG at the surface. (b) Energy band schematic across the central region of the IPGFET. A positive $U_{GS}$ increases the depletion region (green), reducing the effective width $w_{eff}$ of the conductive channel.(c) Schematic of a diamond IPGFET with two curved side gates. (d+e) SEM images of diamond IPGFET structures with two curved side gates (d) as well as a T-shaped single side gate (e). Scale bars are \SI{200}{\nano\metre}. (f) Transistor characteristics of a diamond IPGFET. (g) Dependence of conductance $G$ on the side gate voltage $U_{GS}$ (\SI{50}{\nano\metre} channel width, curved side gates). Colored symbols correspond to the data in (e). The saturation region and linear region are indicated by dashed lines.}
\end{figure}

Bulk diamond, with a band-gap of $E_{\text{gap}} = \SI{5.45}{\electronvolt}$ is considered an insulator or wide bandgap semiconductor. It therefore came as a big surprise when Landstrass and Ravi discovered a high conductivity in undoped, chemical vapor deposited diamond films \cite{Landstrass.1989}. It took years until Maier \textit{et al.} came up with an explanation, the so-called transfer-doping model \cite{Maier.2000}, which is nowadays widely accepted. It ascribes the diamond surface conductivity to the presence of a two-dimensional hole gas (2DHG) at a hydrogen-terminated diamond surface. The 2DHG is established as a result of two necessary conditions: First, a hydrogen-termination of the diamond surface, and second, the presence of atmospheric adsorbates. The C-H dipoles of the H-terminated surface induce a negative electron affinity that effectively shifts the valence band maximum closer to the vacuum level. The atmospheric adsorbates at the diamond surface provide acceptor states with energies $\mu_{ec}$ that are low enough such that electrons can be extracted from the valence band. Thermal equilibrium yields a band bending at the diamond surface with an electrostatic potential well filled by free holes (Figure~\ref{fig:1_IPGFET_RT},~a). In contrast to the H-terminated diamond surface, the energy bands of an O-terminated surface lie around \SI{2.7}{\electronvolt} lower, such that no 2DHG can be established and the surface remains insulating \cite{Maier.2001, Garrido.2008}. Based on this phenomenon, electronic devices such as diamond solution-gated field effect transistors (SGFETs) have been developed \cite{Kawarada.2001, Dankerl.PRL}, where an electrolyte serves as a top gate to modulate the surface conductivity. They have proven to be excellent candidates for the fabrication of biosensors, showing pH \cite{Garrido.2005} and ion sensitivity \cite{Hartl.2007}, as well as being capable of detecting action potentials from cells \cite{Dankerl.2009} or neurotransmitters \cite{Hartl.2004}. In addition, in-plane gated field effect transistors (IPGFETs) have been demonstrated \cite{Wieck.1990}. In these devices charge carriers are laterally depleted from a nanometer-sized H-terminated channel \cite{Garrido.2003, Garrido.2003b}. In the design of diamond-based IPGFETs, thin O-terminated potential barriers separate one or two side gate regions from a conductive H-terminated drain-source channel, as depicted in Figure~\ref{fig:1_IPGFET_RT},~(b). The devices are designed such that a drain-source current through the channel has to pass through a nanometer-sized constriction (Figure~\ref{fig:1_IPGFET_RT},~d,e). The electric field across the O-terminated potential barriers creates a depletion region which reduces the effective width $w_{\text{eff}}$ of the drain-source channel (Figure~\ref{fig:1_IPGFET_RT},~b).

In this letter, we employ in-plane gates to laterally reduce the dimensionality of the 2DHG at the H-terminated diamond surface. We directly prove the two-dimensional nature of the diamond surface conductivity by the observation of a 2D-characteristic gate dependence of the Coulomb blockade effect in the drain-source channel conductance at liquid helium temperature. These effects are the result of zero-dimensional quantum islands, electrostatically generated by the lateral gates in the narrow constriction of the channel.

We used single crystalline, electronic grade diamond substrates (Element Six Ltd.) with an rms surface roughness below \SI{1}{\nano\metre}. The diamond samples were hydrogen-terminated in a microwave-assisted hydrogen plasma as published previously \cite{Dankerl.2009}. Ohmic gold contacts were fabricated by means of photolithography, metal evaporation and selective etching \cite{Dankerl.2009}. O-terminated, non-conductive surface regions were created by selective exposure of the diamond surface to an oxygen plasma. Conventional photolithography and electron-beam lithography were used to realize O-terminated lines with a minimum width of \SI{100}{\nano\metre}, and devices with a minimum channel width of around \SI{50}{\nano\metre}. The scanning electron microscope (SEM) images in Figure~\ref{fig:1_IPGFET_RT},~(d and e) depict two different surface termination patterns in the central region of a device. Two curved side gates (Figure~\ref{fig:1_IPGFET_RT},~e) confine the channel from two sides, or only one side gate is used in a T-shaped design (Figure~\ref{fig:1_IPGFET_RT},~d). The latter provides a minimum channel length which is similar to the width of the O-terminated line. The low-temperature characterization was performed either in a liquid helium bath or in vacuum in a closed cycle cryostat at $T\approx$~\SI{8}{\kelvin}. Lock-in techniques were used for the characterization of the devices. 

We first investigated the diamond in-plane gated field-effect transistors at room temperature in air. Figure~\ref{fig:1_IPGFET_RT},~(f) shows the current-voltage characteristics of a \SI{50}{\nano\metre} wide diamond IPGFET (curved side gates) for gate-source voltages $U_{GS}$ between \SIlist[range-units = single]{-1.0; +2.6}{\volt}. As can be observed, the channel conductivity decreases with more positive gate voltages. Further, the transistor curves exhibit a linear dependence of the drain-source current $I_{DS}$ on the drain-source voltage $U_{DS}$ for low values of $U_{DS}$, and saturation at higher values~\cite{Supplemental.Material}. Figure~\ref{fig:1_IPGFET_RT},~(g) shows the gate-voltage dependence of the channel conductance $G$. The transistor is fully open when $U_{GS} <$~\SI{0}{\volt}, showing a constant conductance. The channel starts to close linearly for more positive voltages until it is closed beyond the threshold voltage $U_{th}$, which is defined as the gate voltage where the extrapolation of the linear current regime equals zero. We have confirmed a linear dependence of $U_{th}$ on the lithographically designed channel width $w_{\text{geo}}$~\cite{Supplemental.Material}.

We can understand these results by using a basic description of the diamond in-plane gates following references \cite{Wieck.1990, Garrido.2003, Garrido.2003b} and considering the energy band schematic in Figure~\ref{fig:1_IPGFET_RT}, (b). At the interface between the oxygen- and hydrogen termination, the energy bands are offset by a built-in potential $e U_{bi}$, which is governed by the different polarities of the surface dipoles in the absence of interfacial charge. Even at $U_{GS}=$~\SI{0}{\volt}, the built-in voltage $U_{bi}$ creates a small depletion region (green shaded areas) in the vicinity of the interface, similar to the case of a pn-junction. The application of a gate voltage across the O-terminated potential barrier modifies the energy bands at the diamond surface, as shown in Figure~\ref{fig:1_IPGFET_RT},~(b). Far away from any interface between the oxygen- and hydrogen termination, the relative position of the valence band edge $E_{\text{VBM}}$ and the Fermi-level $E_F$ stays constant within the H-terminated regions. However, their absolute energy levels are offset by $e U_{GS}$ when a gate voltage is applied across the O-terminated potential barrier. Close to the interface, the gate voltage modifies the energy band bending. As a result, the extension of the depletion region $w_{\text{dep}}^{\text{2D}}$ varies according to
\begin{equation}
	\label{eq:depletion}
	w_{\text{dep}}^{\text{2D}} = \frac{\epsilon \epsilon_0}{e n_h} (U_{bi} \pm U_{GS})
\end{equation}
as described by Petrosyan and Shik for a two-dimensional system \cite{Petrosyan.1989}. 
Here, $\epsilon$ and $\epsilon_0$ are the dielectric constants of diamond and vacuum, respectively, $e$ is the elementary charge, and $n_h$ the carrier concentration (\si{\per\square\centi\metre}) in the 2DHG. The voltage-governed increase of the depletion region (Equation \ref{eq:depletion}) linearly reduces the effective channel width $w_{\text{eff}} = w_{\text{geo}} - 2 \cdot w_{\text{dep}}^{\text{2D}}$. This is directly reflected in the linear decrease of the channel conductance with the gate voltage, as observed experimentally (Figure~\ref{fig:1_IPGFET_RT}, (g)). We can understand $U_{th}$ as the voltage where $w_{\text{eff}} = 0$. In addition, the linearity of $w_{\text{dep}}^{\text{2D}}$ can explain the observed linear dependence of $U_{th}$ on $w_{\text{geo}}$~\cite{Supplemental.Material}. Saturation of the conductance is observed for negative $U_{GS}$, as the channel can not be made wider than its lithographic width ($w_{\text{eff}}=w_{\text{geo}}$). 

\begin{figure}
	\includegraphics{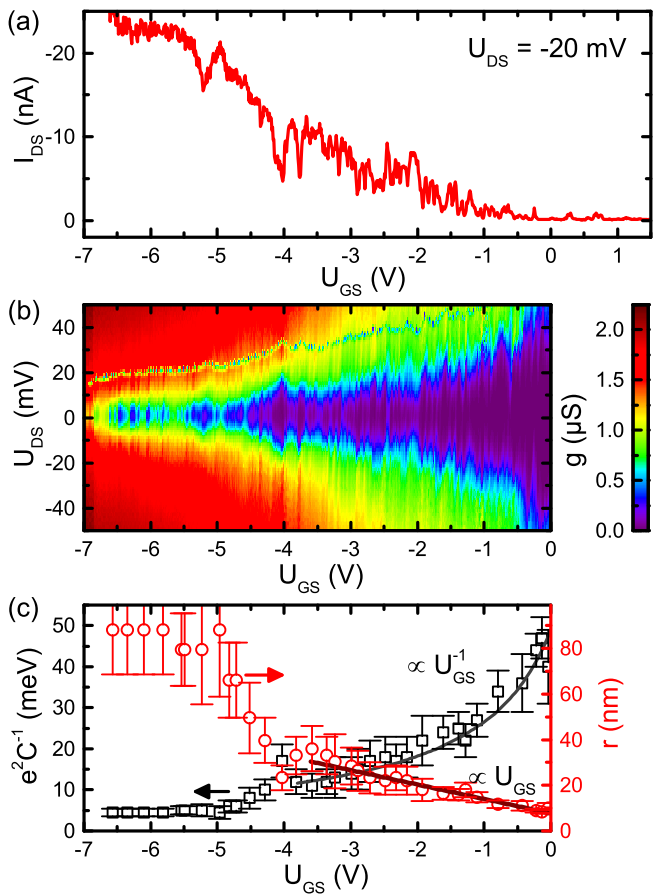}
	\caption[Low-temperature characterization of a diamond IPGFET.]{\label{fig:2_overview_3K} Low-temperature characterization of a diamond IPGFET. (a) Gate-voltage dependence of the drain-source current $I_{DS}$ at $U_{DS}=$~\SI{-20}{\milli\volt}. (b) Color map of the differential conductance $g$ for different gate and drain-source voltages, $U_{GS}$ and $U_{DS}$ (two-sided curved gates, \SI{50}{\nano\metre} width, $T\approx \SI{3}{\kelvin}$). (c) Evaluation of the charging energy $e^2 C^{-1}$ of the Coulomb diamonds in (b). A grey line indicates a $U_{GS}^{-1}$ proportionality. The radius $r$ is calculated assuming a circular disk shape of the quantum dot. From the linear fit, indicated by a dark red line, the 2DHG carrier density can be calculated.}
\end{figure}
    
We further characterized the diamond IPGFETs at liquid helium temperatures to allow quantum effects to become observable. Figure~\ref{fig:2_overview_3K},~(a) shows the gate-voltage dependence of the drain-source current. As shown at room temperature, the channel closes for more positive $U_{GS}$. However, a variety of oscillatory peaks in $I_{DS}$ is visible. For a detailed investigation we measure the differential conductance $g = \partial I_{DS} / \partial U_{DS}$ for different combinations of $U_{DS}$ and $U_{GS}$ (Figure~\ref{fig:2_overview_3K},~(b)).  Unlike at room temperature, $I_{DS}$ does not show an ohmic behavior with $U_{DS}$ ($g(U_{DS}) =$~const.). There exists a central, non-conductive region (purple), which increases in size when the channel closes, i.e. at more positive $U_{GS}$. On the large gate voltage scale \cite{Figure.S2}, the edge of the purple region does not appear smooth but instead Coulomb diamonds  appear, as will be discussed in the following. The green line ($g \approx$~\SI{0.8}{\micro\siemens}) parallel to edge of purple region is a measurement artifact and will be ignored in the following discussion.

We attribute our observations to Coulomb blockade effects in the nanometer-sized central region of the diamond IPGFETs. Coulomb blockade is observed when a small conductive region is separated by tunneling barriers from the drain and source contact \cite{Beenakker.1991}. Due to the small size of the quantum island, it shows discrete charging energies with an energy level spacing of $e^2 C^{-1}$, where $C$ is the capacitance of the island. An electron can only tunnel from the source into the island if the island's energy level is aligned or below the electrochemical potential of the source contact. From the island, it can only tunnel out if the drain level is below the energy level of the Coulomb island. As a result, there exist situations where charge transport across the island is blocked (Coulomb blockade) \cite{Beenakker.1991}. Coulomb diamonds are the graphical 2D representation of these regions, when plotting the differential conductance $g$ versus $U_{DS}$ and $U_{GS}$. 

In order to develop a better understanding and to support our claim, Figure~\ref{fig:2_overview_3K},~(c) shows the calculated charging energy inferred from the extension of the non-conductive region in Figure~\ref{fig:2_overview_3K},~(b). In the case of a 2D conductive layer, the capacitance $C$ of a single Coulomb island with a circular disk shape is related to the islands's radius $r$ via $C = 8 \epsilon \epsilon_0 r$ \cite{Tilke.2001}. A large charging energy of \SI{49}{\milli\volt} is measured at around $U_{GS}=$~\SI{0}{\volt}, which corresponds to an islands radius of about \SI{8}{\nano\metre}. The radius increases gradually as more negative voltages start to open the transistor channel until the tunneling barriers of the island vanish and the channel exhibits an ohmic behavior at $U_{GS}=$~\SI{-7}{\volt}. Between \SIlist{-3.6; -0.4}{\volt} a linear dependence of $r$ on $U_{GS}$ is observed. Since the radius is coupled to the QD charging energy $e^2 C^{-1}$ via the capacitance, $C \propto r$, an inverse proportionality of the charging energy on $U_{GS}$ can be observed in the same region. Such a gate voltage dependence of the charging energy and the radius is characteristic of a 2D conductive channel, as explained below. Let's start by assuming that the island's radius depends directly on the size of the depletion region via $r = r_{\text{max}} - w_{\text{dep}}^{\text{2D}}$. Here, $r_{\text{max}}$ is some maximum value for the radius of the island. In a first approximation, the radius is assumed to decrease by the same amount the depletion region increases in size. Based on this assumption and using Equation (\ref{eq:depletion}), we can now estimate the carrier concentration from the slope of a linear fit of $r$ in the region between \SIlist{-3.6; -0.4}{\volt}, which results in a value around $n_h \approx$~\SI{4E12}{\per\centi\metre\squared}. For an independent confirmation of this value, we have performed transport measurements using Hall bar structures on comparable diamond substrates and with the same fabrication technology as described above. In these experiments~\cite{Supplemental.Material}, a carrier concentration of $n_h=$~\SI{9E12}{\per\centi\metre\squared} could be determined, which compares well to the value inferred from the gate voltage dependence of the Coulomb island radius.

\begin{figure}
	\includegraphics{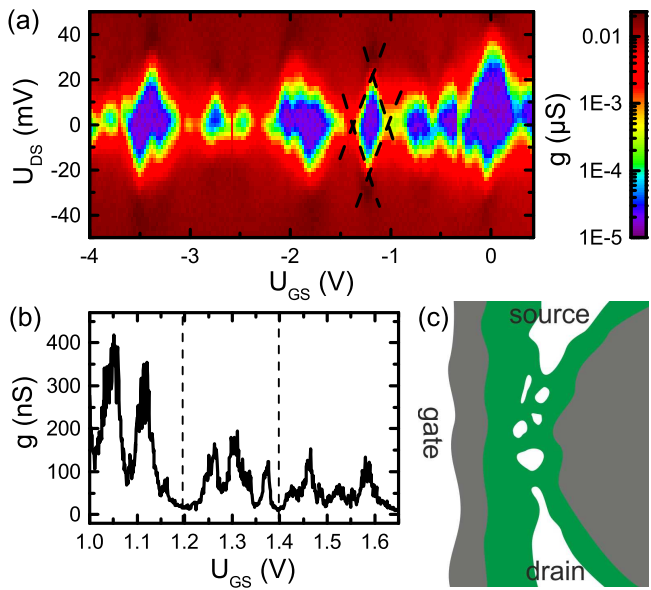}
	\caption[Complex pattern of Coulomb diamonds resulting from coupled multiple quantum dots.]{\label{fig:3_detail_3K} Complex pattern of Coulomb diamonds resulting from coupled multiple quantum dots. (a) Coulomb diamonds in a T-shaped gate structure of \SI{50}{\nano\metre} width at $T =$~\SI{6}{\kelvin}. Dashed, black lines are guides to the eye. (b) Coulomb blockade peaks in a similar device ($U_{DS} =$~\SI{-10}{\milli\volt}), revealing an arrangement of the peaks in three groups with three peaks each. (c) Schematic top-view of the central channel region. The dark gray regions represent the O-terminated areas, whereas the green region sketches the extent of the depletion region. Several conductive islands (white regions) can be formed in the constriction between drain and source due to technology-related imperfections.}
\end{figure}

Figure~\ref{fig:3_detail_3K},~(a) shows a scan similar to Figure~\ref{fig:2_overview_3K},~(b) but for a T-shaped single side gate. Here, individual Coulomb diamonds can be clearly identified, as well as overlapping diamonds. From the height of the individual Coulomb diamond at $U_{GS}=$~\SI{-1.2}{\volt} a capacitance of $C=$~\SI{7.3}{\atto\farad} and a radius of $r=$~\SI{18}{\nano\metre} can be derived. The different slopes of the Coulomb diamond edges indicate an asymmetry of the tunneling barriers between the quantum island and the drain and source contacts \cite{Kaiser.2008}. Figure~\ref{fig:3_detail_3K},~(b) shows a high resolution scan of $g$ at $U_{DS}=$~\SI{-10}{\milli\volt} for a different sample but with the same channel geometry. No regular spacing of Coulomb peaks is observed, as would be expected for a single quantum dot \cite{Kastner.1992, Tilke.2001}. However, the amplitude and spacing of the peaks suggests three groups each with three peaks. We tentatively attribute this observation to the multi-dot splitting which Waugh \textit{et al.} have observed for three electrically coupled quantum dots in series \cite{Waugh.1995}. Figure~\ref{fig:3_detail_3K},~(c) attempts to sketch a more realistic picture of a diamond in-plane gated (T-shaped) structure close to the threshold voltage. Between the depletion zone (green) from the built-in voltage and the depletion zone created by the gate, several conductive islands are localized in the center of the in-plane gate FET structure. The existence of multiple islands of varying shape and size can be understood considering the following aspects. First, the lithographically designed O-terminated regions of the diamond surface (dark grey regions) show a less-than-perfect shape due to irregular broadening of the resist during O-plasma exposure. Second, the surface might have an imperfect hydrogen termination, creating an inhomogeneous potential landscape. Increasing the size of the depletion region via the applied $U_{GS}$ will isolate Coulomb islands step by step from the drain or source regions, while continuously changing the dimension of the island. As a result, not only the electrochemical potentials of the individual islands are changed but also the inter-level spacing of their charging energies. In addition, also the size and height of the tunneling barriers between the islands are influenced and, therefore, the inter-island capacitances. Very similar effects have been observed in graphene nanoribbons by Gallagher \textit{et al.} \cite{Gallagher.2010, Sols.2007}. An irregular arrangement of quantum dots which are interconnected in series and in parallel will yield a complex pattern of Coulomb diamonds \cite{Gallagher.2010}. This can explain the noisy edge of the non-conductive region in Figure~\ref{fig:2_overview_3K},~(b) that results from an overlap of Coulomb diamonds with varying size and shape. Here, the long drain-source channel between the curved side gates allows many conductive islands to form. Only at high negative $U_{GS}$ the channel has opened so much that just one single conductive island remains. This island seems to keep a constant size, as revealed by the constant charging energy observed for large negative voltages in Figure~\ref{fig:2_overview_3K},~(c). In this situation, $U_{GS}$ mostly tunes the electrochemical potential of the dot, leading to the rather periodic Coulomb diamond spacing observed between $U_{GS}=$~\SIlist{-5.8; -6.8}{\volt}. In comparison, the T-shaped single gate structures used to record Figure~\ref{fig:3_detail_3K},~(a) and (b) provide a much shorter channel length. Here, a lower number of conductive islands can be created, which results in more distinct Coulomb blockade effects from only a few islands \cite{Waugh.1995, Gallagher.2010}.
	
In conclusion, we have fabricated in-plane gated FETs on H-terminated diamond with nanometer-sized channel dimensions. Using the lateral control provided by the in-plane gates, it is possible to reduce the effective channel width until a Coulomb island with dimensions down to \SI{10}{\nano\metre} is formed, revealing Coulomb blockade effects at low temperature. We could confirm that the electrostatic control of the island dimensions via the gate voltage shows a dependence characteristic of a 2D carrier density, confirming the two-dimensionality of the diamond's surface conductive channel observed in hydrogenated diamond.

\begin{acknowledgments}
	The authors acknowledge the support of Markus Stallhofer and the financial support of the Graduate School for Complex Interfaces (CompInt), the Nanosystems Inititative Munich (NIM), and the DFG (Diamant Forschergruppe, 1492).
\end{acknowledgments}

\bibliography{lowtemp_IPGFET}

\end{document}